\documentclass{article}

\usepackage[overload]{textcase}
\usepackage{PRIMEarxiv}
\usepackage{tabularx}
\usepackage[table]{xcolor}
\usepackage{makecell}
\usepackage[utf8]{inputenc} 
\usepackage[T1]{fontenc}    
\usepackage{hyperref}       
\usepackage{url}            
\usepackage{booktabs}       
\usepackage{amsfonts}       
\usepackage{nicefrac}       
\usepackage{microtype}      
\usepackage{lipsum}
\usepackage{fancyhdr}       
\usepackage{float}  

\usepackage[utf8]{inputenc}
\usepackage{graphicx}
\usepackage{caption}
\usepackage{subcaption}

\usepackage{graphicx}       
\usepackage{subcaption}
\graphicspath{{media/}}     
\usepackage{caption}
\usepackage{amsmath}
\usepackage{listings}
\usepackage{xcolor}
\title{\MakeLowercase{Deep Learning Inductive Biases for fMRI Time Series Classification during Resting-state and Movie-watching}
}

\author{
  Behdad Khodabandehloo\textsuperscript{a*}, Reza Rajimehr\textsuperscript{b} \\ \\
  \textsuperscript{a}School of Electrical and Computer Engineering, University of Tehran, Tehran, Iran\\
  \textsuperscript{b}School of Cognitive Sciences, Institute for Research in Fundamental Sciences (IPM), Tehran, Iran \\ \\
  \textsuperscript{*}Corresponding Author \\
  Email address: b.khodabandehloo@ut.ac.ir \\
}

\begin{document}
\maketitle

\begin{abstract}

Deep learning has advanced fMRI analysis, yet it remains unclear which architectural inductive biases are most effective at capturing functional patterns in human brain activity. This issue is particularly important in small-sample settings, as most datasets fall into this category. We compare models with three major inductive biases in deep learning including convolutional neural networks (CNNs), long short-term memory networks (LSTMs), and Transformers for the task of biological sex classification. These models are evaluated within a unified pipeline using parcellated multivariate fMRI time series from the Human Connectome Project (HCP) 7-Tesla cohort, which includes four resting-state runs and four movie-watching task runs. We assess performance on Whole-brain, subcortex, and 12 functional networks. CNNs consistently achieved the highest discrimination for sex classification in both resting-state and movie-watching, while LSTM and Transformer models underperformed. Network-resolved analyses indicated that the Whole-brain, Default Mode, Cingulo-Opercular, Dorsal Attention, and Frontoparietal networks were the most discriminative. These results were largely similar between resting-state and movie-watching. Our findings indicate that, at this dataset size, discriminative information is carried by local spatial patterns and inter-regional dependencies, favoring convolutional inductive bias. Our study provides insights for selecting deep learning architectures for fMRI time series classification.

\end{abstract}


\section{Introduction}


Functional MRI (fMRI) provides high-dimensional, multivariate time series that capture brain activity at rest and during stimulation, offering a validation framework for data-driven methods  \cite{vanderwal2019movies,finn2021movie,saarimaki2021naturalistic}. In recent years, deep learning has emerged as a powerful approach for modeling such signals. However, its success depends on the inductive bias built into the chosen architecture, that is, the assumptions a model makes about the structure of the data  \cite{hullermeier2013inductive}. Understanding which inductive biases align best with fMRI time series is therefore essential for designing models that can reliably extract discriminative patterns from this data modality \cite{leming2021deep,weis2020sex,ryali2024deep}. In this regard, sex represents one of the robust axes of inter‑individual variability, influencing cognition, vulnerability to psychiatric and neuro‑degenerative disorders, and treatment response \cite{shansky2016considering, gobinath2017sex, clayton2014policy, wingo2023sex}. Accordingly, sex offers a biologically grounded, widely available label that can serve both as a target for discovery and as a principled benchmark for evaluating whether model architectures capture functional patterns. 


Inductive bias is the set of assumptions that leads a model to favor some hypotheses over others by constraining the hypothesis space. Without such guidance, a model may struggle to discern patterns and become prone to overfitting or underfitting \cite{Zhang2017,LeCun2015, mitchell1980need,WolpertMacready1997, conwell2024large}. Inductive biases can be explicit, predefined rules or constraints, or implicit, arising from architectural choices and hyperparameters \cite{Neyshabur2017,Soudry2018,LeCun2015}. They strongly influence generalization, particularly in low–sample-size settings such as medical datasets \cite{Litjens2017}. When training data are scarce, stronger inductive biases can help the model generalize by steering it toward plausible solutions. Conversely, when data are abundant, weaker biases may be preferable to avoid unnecessary constraints on the search over hypotheses \cite{Dosovitskiy2021,Touvron2021}. Different deep learning architectures embody different inductive biases \cite{Battaglia2018,Raghu2021}. Convolutional neural networks (CNNs) embody a strong locality bias, meaning they assume that nearby elements are correlated and that patterns can recur at different points in time. This makes them particularly effective at recognizing short, repeating patterns, as convolutions are well-suited for capturing short-term temporal patterns. Through weight sharing and limited receptive fields, CNNs achieve approximate translation equivariance and naturally favor short-range temporal structure when applied to one-dimensional time series data \cite{LeCun2015,Cohen2016,bai2018empirical}. Long Short-Term Memory networks (LSTMs) encode an autoregressive memory over ordered observations, biasing the model toward learning both short and long-range temporal dependencies. Their gating mechanisms regulate how much information is remembered or forgotten, allowing them to adapt flexibly to variable timescales when temporal order is crucial. This can be powerful when temporal order is crucial but may be parameter-intensive and data-hungry \cite{Hochreiter1997,Bengio1994, lecun2015deep}. Transformer encoders rely on self-attention, giving them an inductive bias toward modeling relationships between all elements of a sequence regardless of their distance. They do not inherently favor locality or order, but use positional encodings to inject temporal structure when needed. This flexibility allows them to capture global dependencies and contextual interactions, often excelling at tasks where distant or non-local relationships matter. However, the absence of strong built-in temporal or locality priors makes them more data-hungry, since they must learn these structures from scratch rather than relying on architectural assumptions \cite{Vaswani2017,Dosovitskiy2021,Raghu2021, dosovitskiy2020image}. In fMRI specifically, adding domain structure to attention mechanism \cite{bedel2023bolt} can partially compensate for this weak bias, while recent benchmarks still show that simple, strong-bias baselines can be difficult to beat when data are limited \cite{popov2024simple}. This perspective motivates our comparison: if sex-related information in fMRI is dominated by local patterns, CNNs' inductive bias should provide higher classification performance. Conversely, if long-range ordering or global interactions are critical, LSTMs or Transformers should be leveraged.


Prior work on deep learning models for fMRI ranges from static connectivity baselines to spatiotemporal architectures, with recent studies showing that simple architectures can achieve performance comparable to more complex ones. Early work showed that static functional connectivity features alone enable reliable sex prediction from resting-state fMRI, with classical classifiers achieving area under the receiver operating characteristic curve (AUROC) values of around 90\% on the HCP dataset \cite{zhang2018functional, weis2020sex}. Building on this, subsequent research moved beyond static connectivity and explored spatiotemporal deep learning approaches, such as a CNN–LSTM framework trained on dynamic connectivity sequences from the HCP dataset \cite{fan2020deep}.  More recently, stochastic-encoded CNNs have been applied to sex classification using both resting-state and task fMRI, achieving nearly 95\% accuracy across rest and multiple tasks in the UK Biobank cohort \cite{leming2021deep}. A spatiotemporal deep neural network based on 1D CNNs uncovered replicable, behaviorally relevant sex differences and demonstrated strong generalization across independent studies while maintaining around 90\% accuracy \cite{ryali2024deep}. Yet, a recent benchmark revealed that a surprisingly simple mean-time-series MLP can outperform more sophisticated architectures on several fMRI classification tasks, underscoring the central role of architectural inductive bias \cite{popov2024simple}. For individual recognition from resting-state fMRI, a convolutional-RNN was shown to outperform a vanilla RNN, an advantage attributed to its superior ability to extract local features between neighboring ROIs \cite{wang2019application}. In state classification, a recent study reported slightly better overall performance for a 1D-CNN than a BiLSTM \cite{kucukosmanoglu2025influence}.


\textcolor{black}{Despite substantial progress, the field still lacks a systematic assessment of how distinct inductive biases, local convolutional filtering, recurrent sequence modeling and global self-attention affect fMRI-based sex classification, especially under a unified pipeline spanning both resting-state and task fMRI data.} Most prior work introduces a single architecture or compares only variants within one family, limiting cross-family analysis \cite{zhang2018functional, leming2021deep}. Even recent benchmarks \cite{popov2024simple} primarily contrast a mean-MLP with a small set of deep neural network models on resting-state data, without isolating the contribution of architectural bias itself. Meanwhile, previous studies that mix multiple components \cite{ryali2024deep} make it difficult to attribute gains to a single inductive bias. A unified comparison on identical parcellated inputs, evaluated consistently across various functional networks, remains absent, leaving open which inductive biases best align with sex-related functional activity.


\textcolor{black}{We performed a comparison of four architectures, 1D CNN, LSTM, LSTM-CNN, and Transformer, trained on parcellated multivariate fMRI time series from the HCP-7T cohort (resting-state and movie-watching task fMRI) and the Glasser parcellation \cite{glasser2016multi}.} Using a unified training and evaluation pipeline, we analyzed Whole-brain, subcortex and 12 functional networks (Language, Orbito-Affective, Posterior Multimodal, Primary Visual, Secondary Visual, Somatomotor, Ventral Multimodal, Cingulo-Opercular, Default Mode, Dorsal Attention, Frontoparietal) \cite{ji2019mapping}. We also conducted additional experiments to further investigate the intrinsic characteristics of fMRI data, aiming to determine whether models rely more on sequence structure or on stationary and spatial patterns. Our results showed that CNNs consistently achieved the highest AUC-ROC for sex classification in both resting-state and movie-watching, whereas LSTM and Transformer did not perform effectively given the available sample size. Network-level analyses identified specific functional networks that were more discriminative, and ablation studies further examined the importance of spatial and temporal patterns. Together, these findings provide a comprehensive comparison of inductive biases across CNNs, LSTMs, LSTM-CNNs, and Transformers, offering practical guidance for model selection in fMRI time series classification.

\section{Materials and Methods}

\subsection{Dataset}


\textcolor{black}{We conducted the analysis using four independent runs of 7T resting-state fMRI data, comprising approximately 180 subjects, and four independent runs of 7T movie-watching task fMRI data, also involving about 180 subjects.} We obtained Both datasets from the HCP-7T cohort.  Comprehensive information regarding the HCP dataset is already available in prior publications \cite{van2012human, glasser2013minimal, smith2013resting}. We employed the HCP functional pipeline for data processing based on methods reported in previous publications \cite{glasser2013minimal}. The primary procedures included: initial spatial processing in both volumetric and gray-ordinate formats, which involves representing brain locations as surface vertices. This was followed by applying a weak high-pass temporal filter with a full width at half maximum greater than 2000s to both data types for the elimination of slow drifts. For the volumetric data, we utilized the MELODIC ICA, and artifacts were identified via FIX. To eliminate artifacts and motion influences, we performed regression using several parameters: the six rigid-body parameter time series, their temporal derivatives looking backward, and the squared values of all 12 regressors, applied to both volumetric and gray ordinate data. For every participant, we extracted regional mean time series using the 379-region Glasser atlas \cite{glasser2016multi} and Z-scored them.

\begin{figure}[ht]
    \centering
    \includegraphics[width=1.0\linewidth]{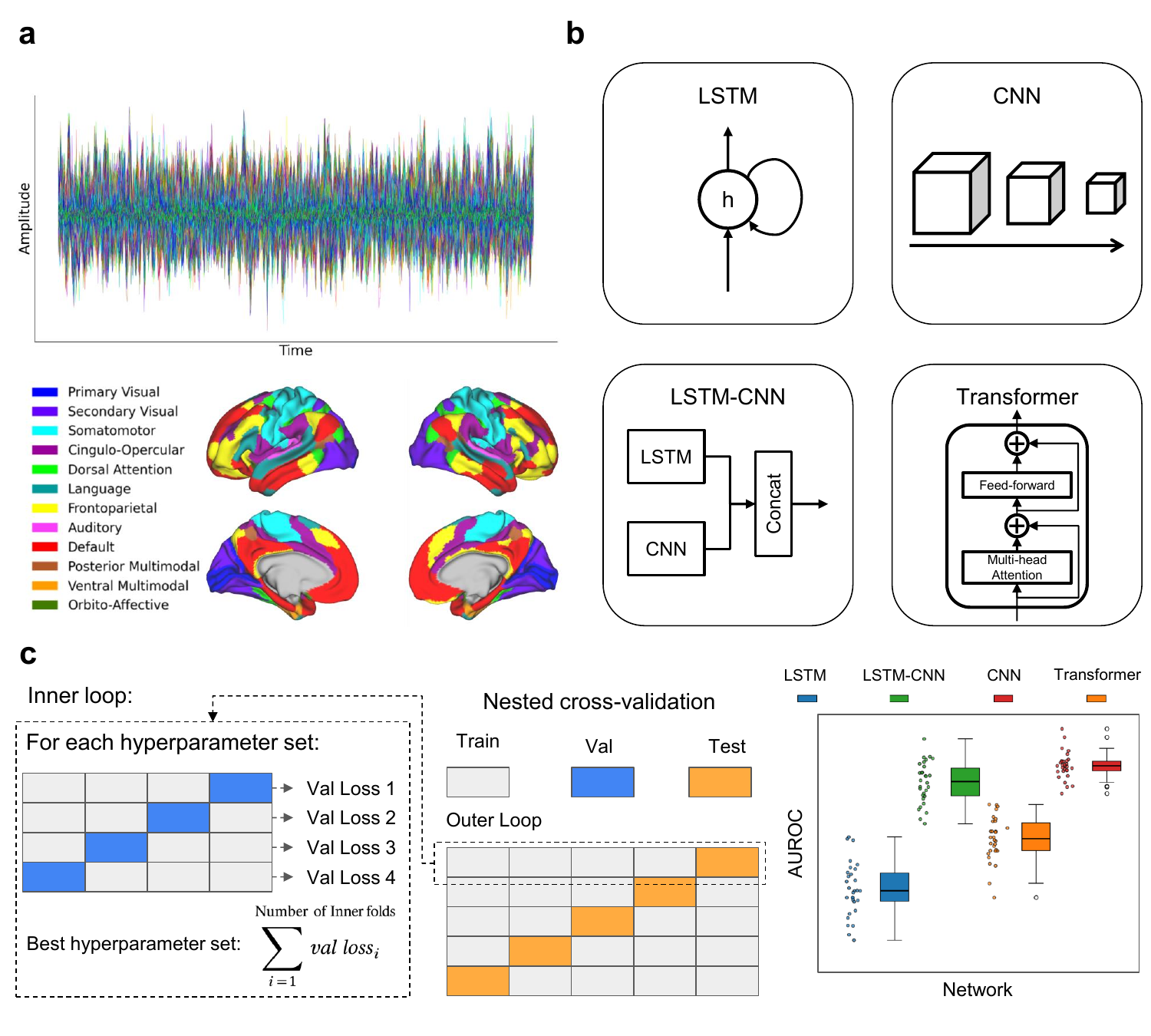}
    \caption{The overall pipeline of our study \textbf{a)} The input to our classification pipeline consists of preprocessed resting-state and movie-watching multivariate fMRI time series from the HCP-7T cohort. We use 12 functional networks, along with Whole-brain and subcortex, as inputs to the models. \textbf{b)} We leverage three major categories of deep learning models, CNNs, LSTMs, and Transformers, each offering distinct inductive biases. \textbf{c)} We assess model performance through an extensive evaluation pipeline based on nested cross-validation. We perform hyperparameter optimization in the inner loop, while we obtain final test results from the held-out test set. We further obtain results across independent runs and random seeds for each model.}
    \label{fig:fig1}
\end{figure}


\textcolor{black}{We used 12 functional networks \cite{ji2019mapping}, together with the Whole-brain and subcortex, for the classification task in both datasets.} The 12 networks include Auditory, Cingulo-Opercular, Default Mode, Dorsal Attention, Frontoparietal, Language, Orbito-Affective, Posterior Multimodal, Primary Visual, Secondary Visual, Somatomotor, and Ventral Multimodal. Each sample therefore consists of a multivariate time series $\mathbf{X} \in \mathbb{R}^{C \times T}$, with $C$ channels and approximately $T = 900$ sequential time points (TRs $=$ 1s). The class labels $\mathbf{y}$ represent the participant’s sex.

\subsection{Multivariate time series classification models}


\textcolor{black}{Selecting a deep learning architecture whose inductive bias align with the inherent characteristics of multivariate fMRI time series is crucial for achieving accurate classification.} In this study, we compare three families of deep learning models, namely Convolutional Neural Networks (CNNs), Long Short-Term Memory networks (LSTMs), and Transformers, to determine which inductive biases are most suitable for learning representations from multivariate fMRI time series for the task of sex classification. Each architecture emphasizes different inductive biases: CNNs are effective at capturing local and spatial patterns, LSTMs are effective in modeling sequential temporal dependencies, and Transformers leverage self-attention to integrate information globally. By training and evaluating these models on the same dataset, we assess how well their respective biases capture the information inherent in fMRI time series, ultimately identifying the approach that yields the most informative representations and, in turn, the highest discrimination capabilities for sex classification using multivariate fMRI time series.


\textcolor{black}{In our convolutional model, we integrate one‑dimensional (1D) convolutions with channel attention to capture the temporal and inter‑regional structure of multivariate fMRI signals.} The CNN we use in this study consists of 1D convolutions. This model is applied along the temporal dimension of fMRI time series. Since we apply 1D kernels along the temporal dimension, the network learns local temporal and inter-channel dependencies, achieves translation invariance in time. Our architecture consists of three sequential 1D convolutional blocks. This builds temporal hierarchies through successive layers. Each block is followed by a squeeze-and-excitation channel-attention module \cite{hu2018squeeze}, and after the final block, global average pooling is applied across time to produce a final embedding. We feed the resulting embedding into a fully connected layer for classification, while an optional 1D max pooling layer after each convolutional block allows us to examine the effect of reduced temporal resolution. We have adapted this architecture from the CNN branch of the multivariate LSTM-FCN framework proposed by \cite{karim2019multivariate}, with the recurrent component originally introduced in \cite{hochreiter1997long} omitted.


\textcolor{black}{To exploit temporal dependencies in fMRI time series, we employ LSTM models \cite{hochreiter1997long} whose gating mechanisms preserve information across short- and long-range intervals.} The model processes the sequence with a configurable stack of unidirectional or bidirectional LSTM layers. To summarize the sequence for classification, we extract the final hidden state from the last time step (or, in the bidirectional case, concatenates the last hidden state from the forward pass with the first hidden state from the backward pass). We pass this final representation through a feedforward classification head. Configurable hyperparameters are the number of layers, hidden size, dropout rate, and whether the LSTM is bidirectional.


\textcolor{black}{To exploit both temporal dependencies and local patterns in multivariate fMRI data, we adopt the Multivariate LSTM‑CNN architecture \cite{karim2019multivariate}.} This model is an extension of LSTM-CNN model in \cite{karim2017lstm} by adding squeeze and excitation blocks to the CNN part of that model. This hybrid model combines an LSTM branch with a CNN branch. After extracting features via both branches, the final representations were concatenated and passed through a shared fully connected classification head. 


\textcolor{black}{Leveraging the global‐context capabilities of self‑attention, we adapt a Transformer architecture to classify multivariate fMRI time series.} We employ a Transformer-based neural network designed for classification of fMRI time series data. Transformer models leverage self-attention mechanisms \cite{vaswani2017attention}, allowing them to model global dependencies in a sequence more directly and flexibly than traditional recurrent architectures such as LSTMs and RNNs. The input features are first linearly projected into a hidden representation suitable for the multi-head self-attention mechanism. For this we use a sinusoidal positional encoding. The encoded sequence is processed by a stack of Transformer encoder layers, parameterized by configurable hyperparameters such as number of layers, number of attention heads, and hidden dimension size. After Transformer encoding, we extract the representation corresponding to the first time step as a global summary, which is passed through a feedforward classification head. We have adapted the model from the architecture proposed in \cite{popov2024simple}.

\subsection{Training and Evaluation}


\textcolor{black}{ To ensure robust and reliable model evaluation, we implemented a comprehensive training and validation framework based on nested cross-validation.} We apply this pipeline consistently to all deep learning architectures used. The inner loop employs 4-fold cross-validation to optimize hyperparameters using the training and validation folds, while the outer loop performs 5-fold cross-validation to assess the generalization performance of the tuned models on unseen test data (Figure \ref{fig:fig1}, c). After the inner-loop optimization, we retrain the model on the full training set (excluding the outer test fold) using the best hyperparameters, and its performance is then evaluated on the outer test fold. This process ensures that the outer test data remain completely isolated from both training and hyperparameter selection, providing an unbiased estimate of model performance and reducing the risk of overfitting during model selection and evaluation. The inner loop’s hyperparameter optimization process is critical to this framework and is conducted systematically to ensure optimal model performance. We repeated the process of training and evaluation using 5 different seeds. 


\textcolor{black}{We employed a systematic approach to hyperparameter optimization within the inner loop of the nested cross-validation framework using Optuna \cite{akiba2019optuna}, an optimization library based on Bayesian optimization with a Tree-structured Parzen Estimator.} The objective was to minimize the average validation loss across the inner folds. For each hyperparameter configuration, validation losses were computed across all folds, and the optimal set was selected based on the lowest total validation loss. Each Optuna study was run with up to 50 trials to efficiently explore the hyperparameter space. 



\textcolor{black}{We train all models using a consistent optimization strategy designed to promote stable convergence and generalization.} We used the Adam optimizer with a weight decay of $1 \times 10^{-4}$ to regularize model weights. We employed a step-based learning rate scheduler (step\_size=10, gamma=0.1) to reduce the learning rate during training. For recurrent architectures, we used gradient clipping technique ($\|g\|_2 \leq 1.0$) to prevent gradient explosion and improve stability. We chose the batch size of 16 for both the inner and outer cross-validation loops, each running for 100 epochs. For each outer fold, we used the model corresponding to the best epoch (based on validation performance) for final evaluation. We assessed the model performance using AUROC on the held-out test fold. AUROC is widely used because it measures a model’s ability to distinguish between classes across all thresholds, making it robust to class imbalance and independent of any specific decision threshold.

\section{Results}


\begin{figure}[!ht]
    \centering
    \includegraphics[width=1.0\linewidth]{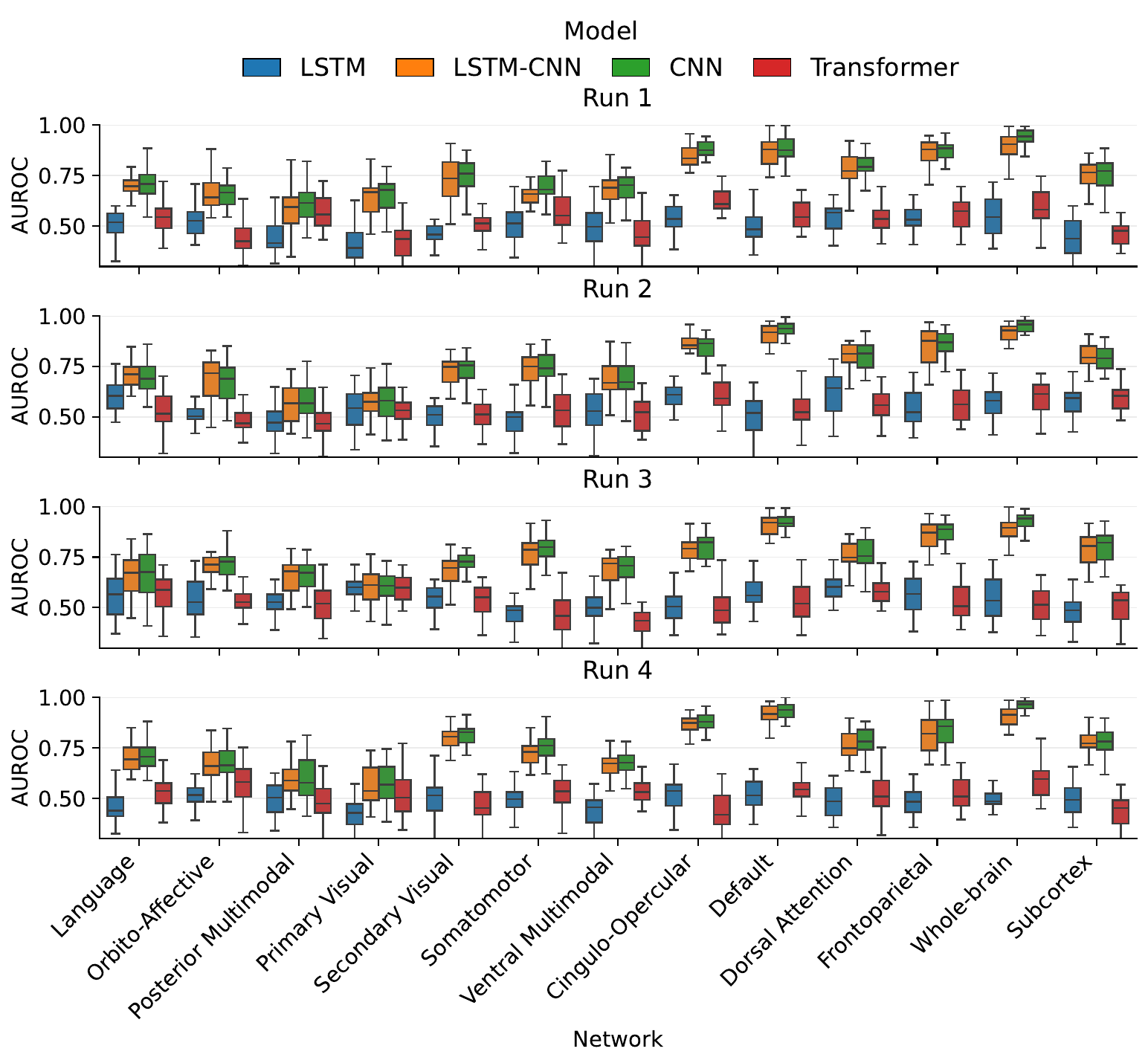}
    \caption{Performance of four models (LSTM, LSTM–CNN, CNN, and Transformer) on resting-state fMRI. Each row corresponds to an independent run.}
    \label{fig:fig_rest_fmri_results}
\end{figure}

In this study, we first compare model performance on sex classification across functional networks and examine which deep neural model’s inductive bias makes it best suited to fMRI time series classification. We then conduct ablation studies to further analyze and interpret the intrinsic characteristics of fMRI data relevant to sex classification.



\begin{figure}[!ht]
    \centering
    \includegraphics[width=1.0\linewidth]{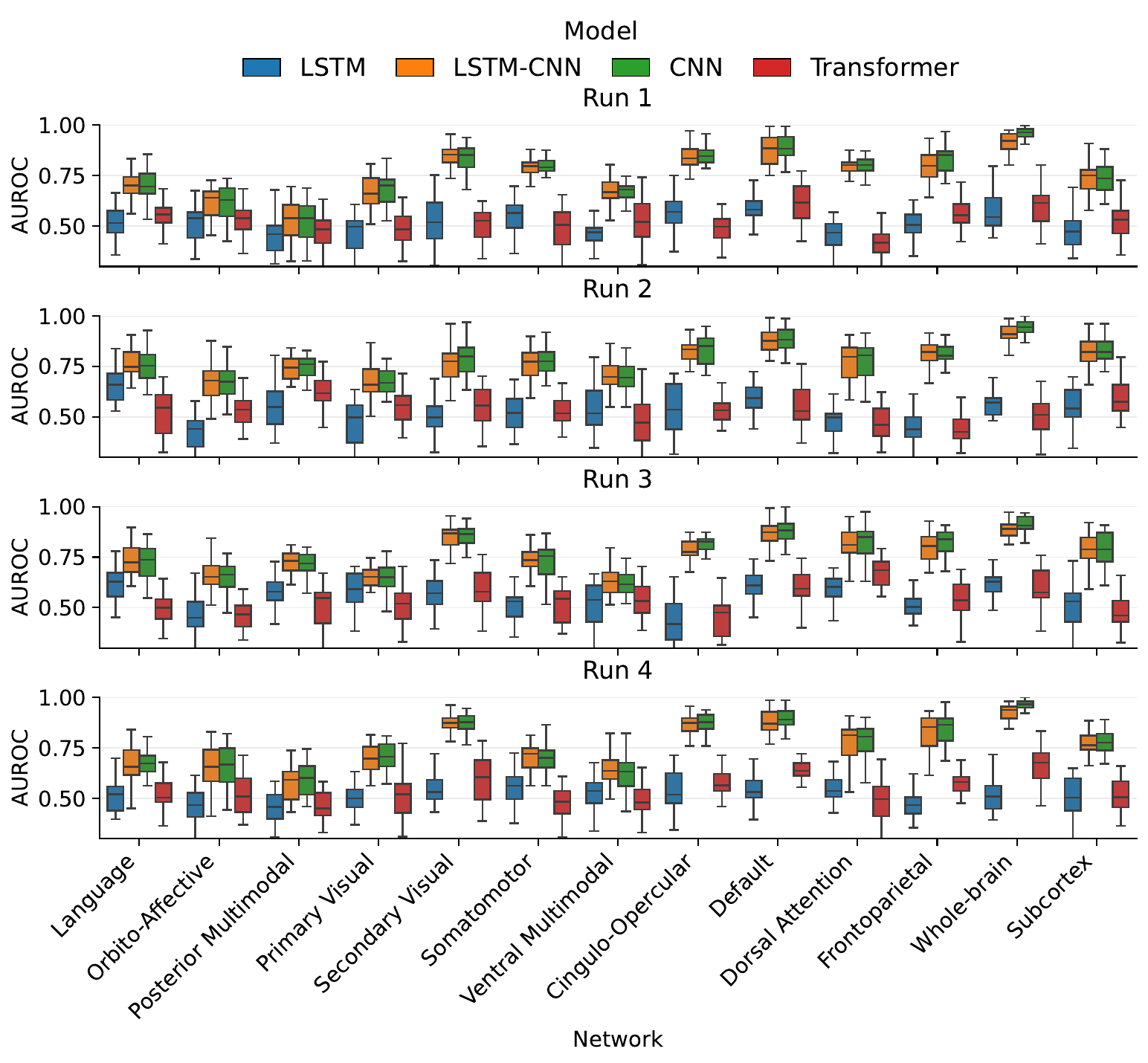}
    \caption{Performance of four models (LSTM, LSTM–CNN, CNN, and Transformer) on movie-watching fMRI. Each row corresponds to an independent run.}
    \label{fig:fig_movie_fmri_results}
\end{figure}

\textcolor{black}{Our first analysis examined which inductive bias was more effective for sex classification using multivariate time series from both resting-state fMRI and movie-watching task fMRI.} Consequently, we compared the performance of four models: CNN, LSTM, LSTM-CNN, and Transformer. We evaluated performance across all previously defined functional networks, as well as the Whole-brain and subcortex, using four independent runs of 7T resting-state and movie-watching fMRI. This independent analysis confirmed the consistency of patterns across runs. The results are shown in Figure~\ref{fig:fig_rest_fmri_results} and Figure~\ref{fig:fig_movie_fmri_results}, with each row corresponding to one independent run. Across both resting-state and movie-watching, functional networks exhibited varying levels of discriminative power for sex classification. CNN and LSTM-CNN models consistently outperformed LSTM and Transformer models, with the latter showing lower discriminative performance. The superiority of CNN-based models was robust across all four runs in both conditions. Although CNN and LSTM-CNN achieved similar performance, CNN outperformed in most cases.


\textcolor{black}{To further compare model inductive biases, we repeated the analysis using univariate averaged fMRI time series.} For each functional network, we computed a univariate time series by averaging across features at each time point, resulting in a univariate time series per network. Then, similar to the multivariate case, we applied each model to these univariate time series (Figure~\ref{fig:univarite_test_task}). The performance of CNN and LSTM–CNN dropped substantially, indicating that cross-ROI dependencies contain important discriminative information, which is largely lost when averaging.

\begin{figure}[!ht]
    \centering
    \includegraphics[width=1.0\linewidth]{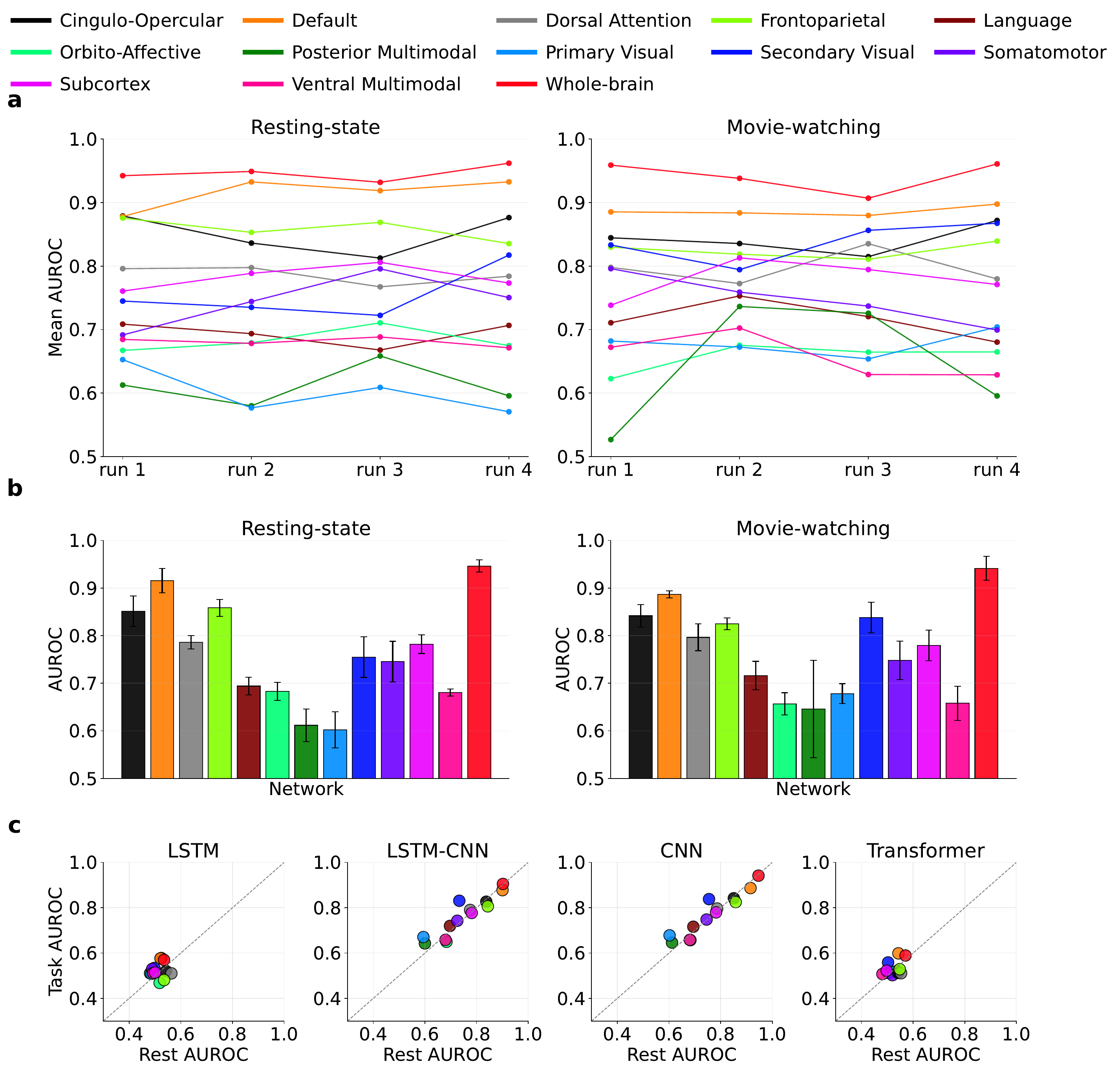}
    \caption{\textbf{a.} Discriminative ability of each network across runs. \textbf{b.} Comparison of discriminative ability across networks. \textbf{c.} Scatter plot of mean AUROC values averaged across all folds and seeds; the x-axis shows resting-state AUROC and the y-axis shows movie-watching AUROC for each functional network.}
    \label{fig:fig_consistency_ab}
\end{figure}
\begin{figure}[!ht]
    \centering
    \includegraphics[width=1.0\linewidth]{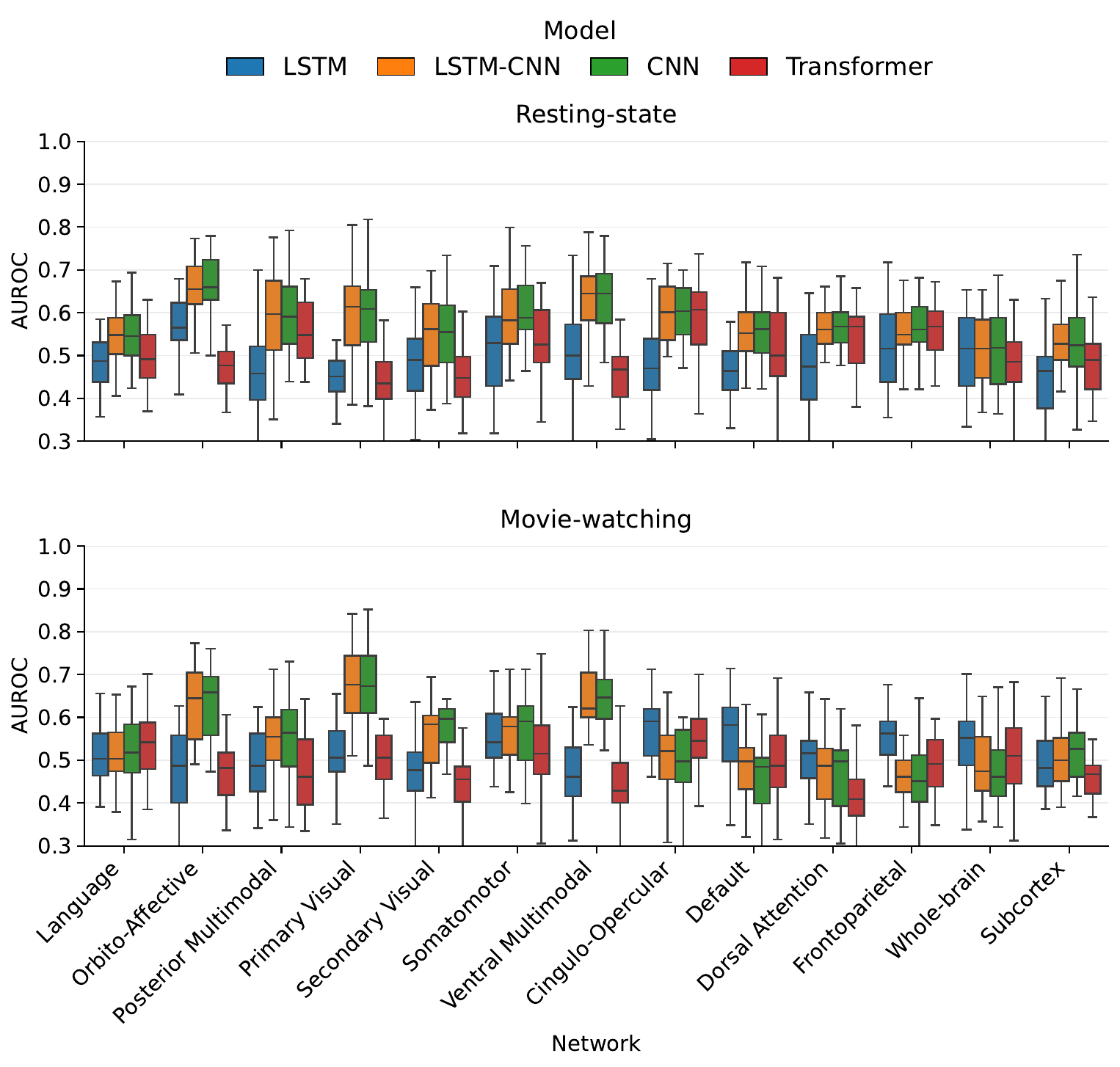}
    \caption{Classification results using univariate time series obtained by averaging across ROIs at each time point. The first row presents resting-state results, while the second row presents movie-watching results.}
    \label{fig:univarite_test_task}
\end{figure}

\textcolor{black} {Our results revealed sex-discriminative functional networks in both resting-state and movie-watching, showing largely similar patterns across conditions.} In resting-state fMRI, the Cingulo-Opercular, Default Mode, Frontoparietal, and Whole-brain networks exhibited high discriminative power, whereas the Orbito-Affective, Posterior Multimodal, Language, and Ventral Multimodal networks provided limited discriminative information (Figure~\ref{fig:fig_consistency_ab}a,b). In movie-watching, we observed similar patterns for the highly discriminative networks, with the Secondary Visual network also appearing highly discriminative, while the Orbito-Affective, Posterior Multimodal, Language, and Ventral Multimodal networks again provided limited discriminative information. Although the Posterior Multimodal network exhibited low variance across runs in resting-state, it showed high variance during movie-watching. A scatter plot of mean AUROC values for resting-state versus movie-watching further revealed a strong correlation in network-level patterns (Figure~\ref{fig:fig_consistency_ab}c). Overall, the Whole-brain exhibited the strongest discriminative power, with the highest mean ROC–AUC and the lowest standard deviation.

\begin{figure}[!ht]
    \centering
    \includegraphics[width=1.0\linewidth]{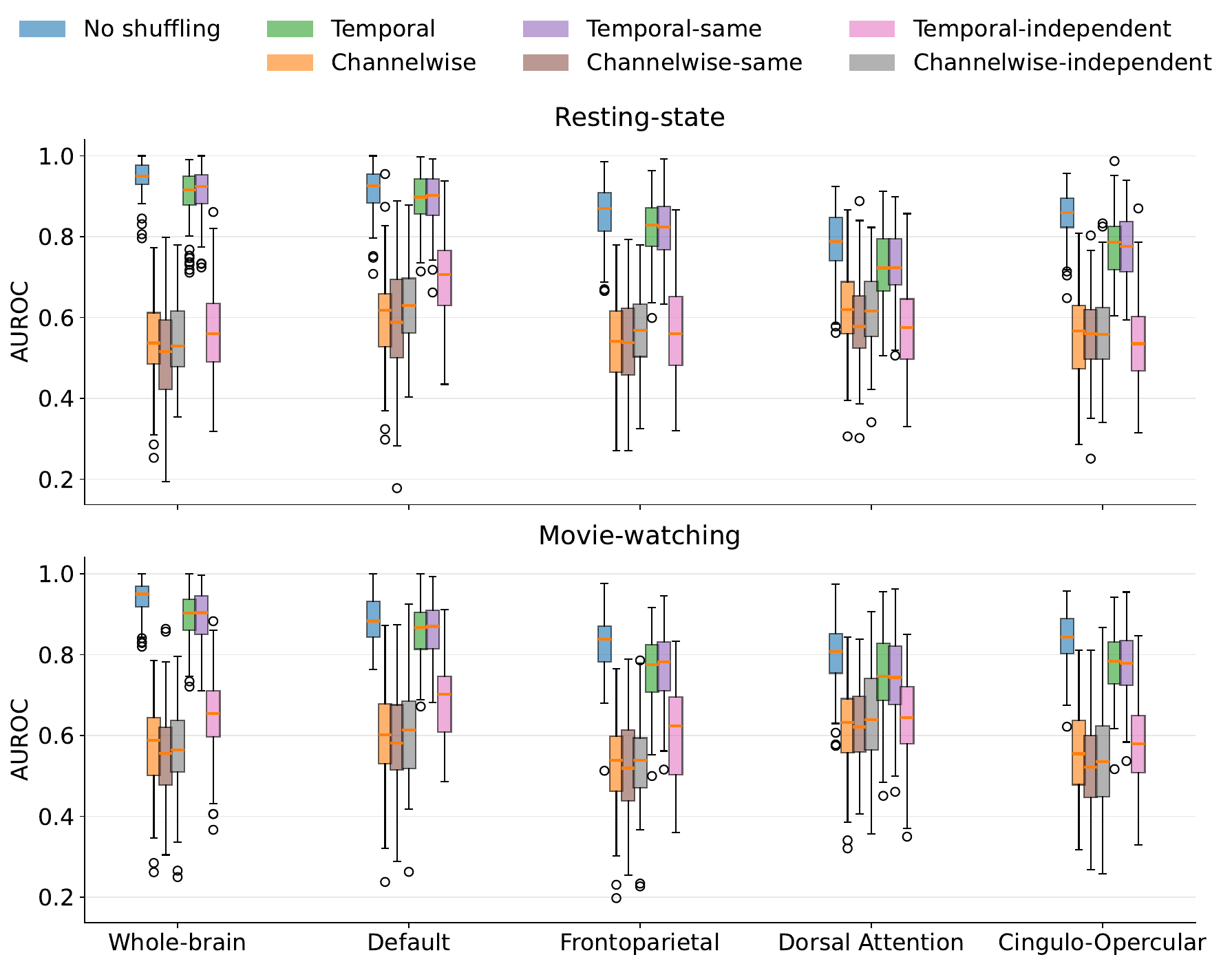}
    \caption{Classification results for different shuffling approaches. The first row shows resting-state results, and the second row shows movie-watching results. Two categories of shuffling were applied: channelwise, which permutes data across the channel dimension, and temporal, which permutes data along the time dimension.}
    \label{fig:fig_shuffling}
\end{figure}

\begin{figure}[!ht]
    \centering
    \includegraphics[width=1.0\linewidth]{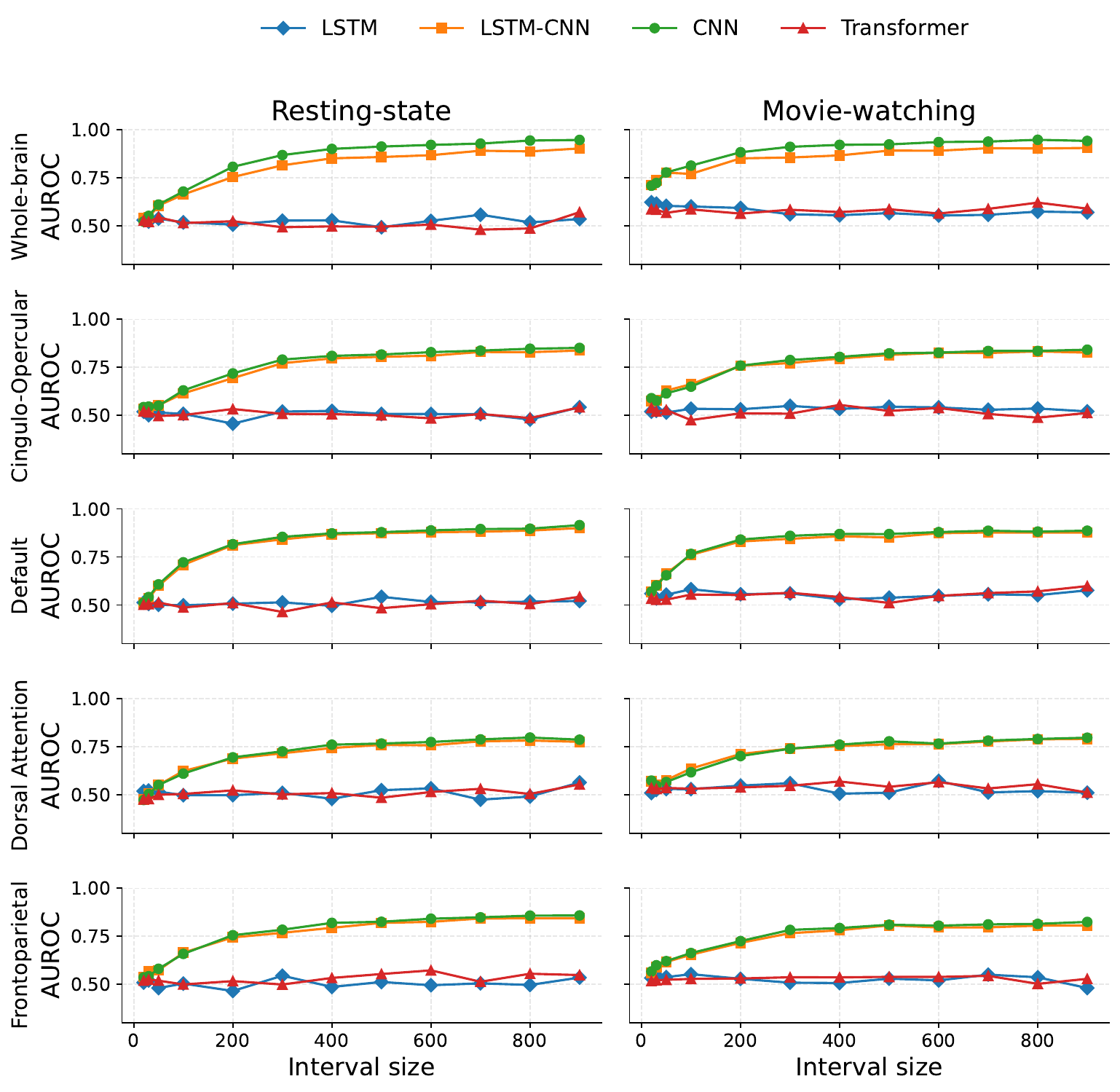}
    \caption{Effect of fMRI duration on classification performance of CNN model in both resting-state and movie-watching.}
    \label{fig:interval_rest_task}
\end{figure}

\begin{figure}[!ht]
    \centering
    \includegraphics[width=1.0\linewidth]{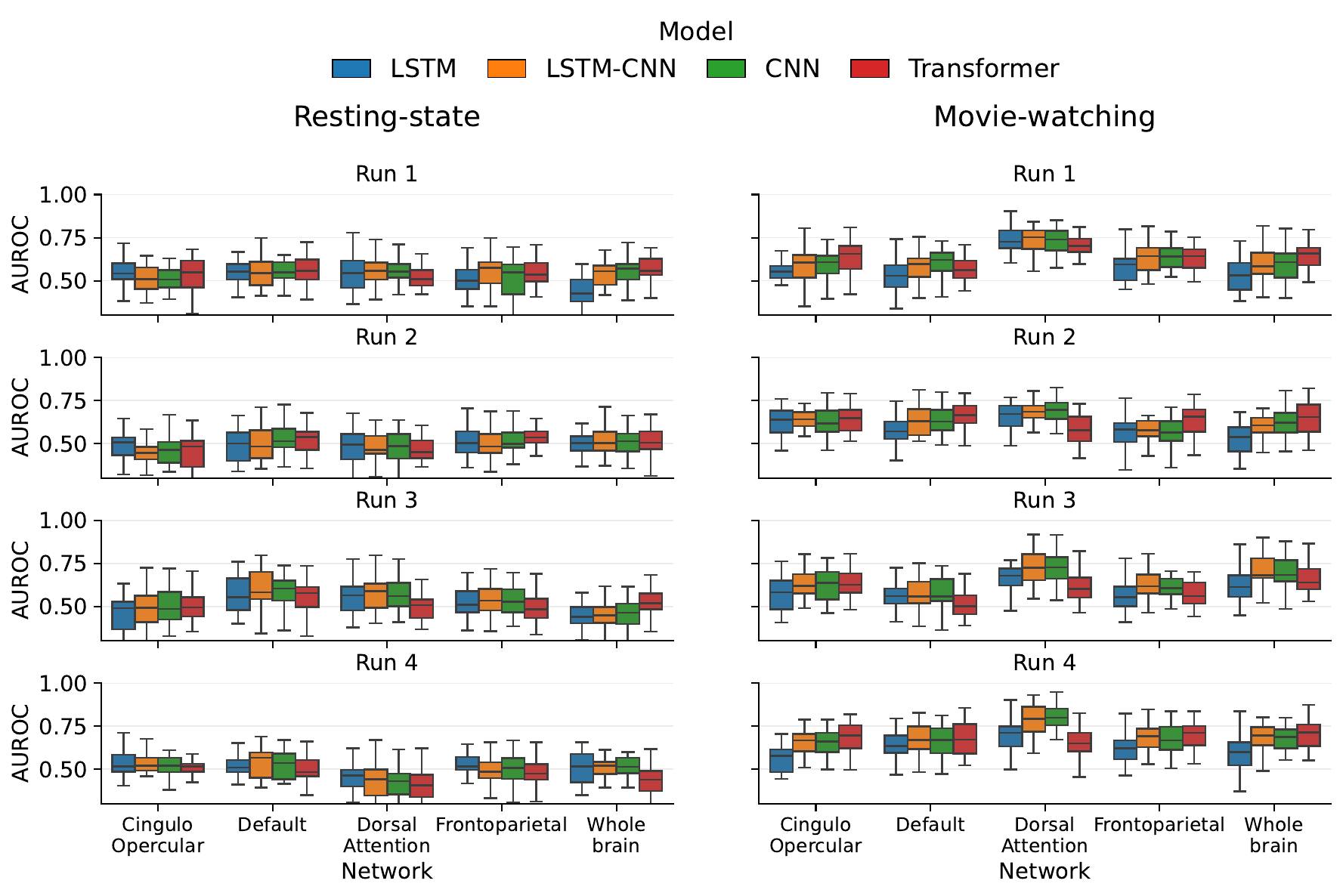}
    \caption{Classification results of the CNN model when trained across the channel dimension instead of the temporal dimension.}
    \label{fig:box_plot_across_channels}
\end{figure}


\textcolor{black}{Using channelwise and temporal shuffles, we further validated that spatial, local and stationary information, not temporal order, provided the dominant discriminative information in fMRI in both resting-state and movie-watching.} We trained models on shuffled multivariate time series and evaluated them on unshuffled validation and test sets, using two shuffling approaches (channelwise and temporal), each with three variants. In the channelwise approach, we permuted ROI positions using three methods: (1) channelwise-same, the same channel permutation for all subjects; (2) channelwise, a subject-specific channel permutation; and (3) channelwise-independent, a subject- and time-point-specific channel permutation, with a new permutation at every time step within each subject. In the temporal approach, we permuted time points using three methods: (4) temporal-same, where the same temporal permutation was applied to all subjects; (5) temporal, where each subject's time series modified using a unique temporal permutation; and (6) temporal-independent, where a separate temporal shuffle was applied independently to each channel within each subject. Channelwise shuffling reduced AUROC to near-chance levels in both resting-state and movie-watching data for all three variants (Figure~\ref{fig:fig_shuffling}). Temporal shuffling caused only marginal decreases in AUROC for temporal-same and temporal, but a substantial drop for temporal-independent (Figure~\ref{fig:fig_shuffling}), indicating that temporal order contributes but is not essential unless cross-ROI alignment is disrupted. These patterns held for both resting-state and movie-watching, and even in movie-watching, despite its sequential structure, spatial information proved more informative than temporal order.


\textcolor{black}{We examined the impact of fMRI signal duration on sex classification performance in both resting-state and movie-watching.} To do so, we randomly sampled time windows ranging from 20 to 900 seconds (20, 50, 100, 200, ..., 900) and evaluated model performance at each duration (Figure~\ref{fig:interval_rest_task}). Overall, longer fMRI durations improved sex classification in both conditions, whereas for short intervals ($<$ 100 s), performance remained below 0.8. During movie-watching, high accuracy was achieved with shorter intervals than in resting-state. Performance generally increased with interval length but plateaued around 600 seconds, with little change beyond this point in the movie-watching. We performed this analysis for the Whole-brain as well as the Frontoparietal, Dorsal Attention, Default Mode, and Cingulo-Opercular networks.


\textcolor{black}{To further assess inherent characteristics of fMRI data for classification, we examined the performance of CNN model when applied across channels rather than time.} In all previous experiments, CNNs were trained along the temporal dimension, but here we repeated the analysis with CNNs applied across channels instead. We focused on five highly discriminative functional networks and found that performance decreased in both resting-state and movie-watching. However, during movie-watching, some discriminative information remained, whereas in resting-state it was largely diminished. In particular, the Dorsal Attention network retained strong discriminative ability in the movie-watching. These results suggested that temporal dependencies may operate across channels during movie-watching (Figure~\ref{fig:box_plot_across_channels}).

\section{Discussion}

This study provides a comparison of deep learning inductive biases for classifying biological sex from multivariate fMRI time series in both resting-state and movie-watching. Across four independent runs in each case, 1D convolutional models consistently outperformed recurrent and attention-based alternatives, with LSTM-CNN close behind CNN and LSTM and Transformer near chance. Network-resolved analyses revealed largely similar patterns between resting-state and movie-watching, with strong discriminative ability in the Default Mode, Cingulo-Opercular, Dorsal Attention, Frontoparietal, and Whole-brain networks. Ablation experiments further indicated that cross-ROI dependencies carries the dominant discriminative information, with temporal order providing less information. Together, these observations indicate that local, stationary patterns align best with the sex-related information at this sample scale, favoring convolutional inductive bias. 


\textcolor{black}{The superiority of CNNs implies that, for the sex classification application, spatial patterns are more discriminative than temporal or global dependencies, a finding consistent across both resting-state and movie-watching.} The limited improvements from adding recurrence (LSTM–CNN vs. CNN) indicate that learning temporal order contributes little beyond what convolutional layers already capture. Meanwhile, The underperformance of Transformers underscores the need for stronger inductive biases, such as local-window attention, which constrains the model to prioritize nearby features and better capture spatial information in fMRI data. Alternatively, substantially larger sample sizes may be required to fully exploit their capacity. These findings are consistent with prior work showing that models ignoring temporal order can achieve high fMRI classification performance \cite{popov2024simple}, and with earlier methods that successfully employed CNNs to extract informative patterns within more complex architectures \cite{ryali2024deep, esfahani2025cnn, farhan2025enhancing}. The significant performance drop when averaging channels into a univariate time series further underscores that averaging features across ROIs destroys the spatial covariance structure exploited by 1D CNNs.



\textcolor{black}{Channelwise shuffling, which breaks ROI order and cross-ROI covariance, reduces AUROC to chance level. In contrast, shuffling across time has little effect on performance unless cross-ROI alignment is independently randomized at each time point, thereby destroying the local multivariate organization.} This experiment argues that which region carries what matters more than exactly when it happens again favoring convolutional inductive bias over recurrent ones. These shuffling experiments change the data distribution, revealing that the model is sensitive to ROI identity but not to temporal order, which confirms reliance on spatial covariance rather than sequence dynamics \cite{popov2024simple}. In practice, this shows that suitable architectures are those that can capture cross-ROI dependencies. In addition, data augmentations should preserve ROI identities and avoid disrupting spatial consistency in fMRI data.


\textcolor{black}{In both resting-state and movie-watching, the most discriminative patterns came from association networks (Default, Cingulo-Opercular, Dorsal Attention, Frontoparietal) and the Subcortex.} These networks show stable, cross-task signatures linked to cognition, arousal and control, and subcortical modulation. Such patterns may reflect consistent sex-related differences that can be detected even without modeling tasks directly. In comparison, sensory and motor systems contributed the least, which fits with weaker sex differences in lower-level processing \cite{shanmugan2022sex, ryali2024deep, serio2024sex}.




Our results are based on a single cohort (HCP-7T) and a single parcellation (Glasser's parcellation), with a small sample size (approximately 180 per run), which may limit the generalizability of our findings. With a larger sample size, model performance may differ. Moreover, in this study we focused only on sex as the classification label, restricting our conclusions to this phenotype. Future work should examine larger and more diverse datasets, evaluate generalization across cohorts and parcellations, and extend beyond binary sex classification to other phenotypes, while maintaining the unified inductive bias comparisons demonstrated here. Our findings also suggest differences between movie-watching and resting-state data, particularly when CNNs were trained across channels, an observation that requires further investigation in future studies. Building on our findings, future models should aim not only to improve performance but also to achieve comparable accuracy using shorter fMRI intervals.

\section{Conclusion}

We conducted a unified comparison of deep learning inductive biases for fMRI time series classification, applied to sex classification from resting-state and movie-watching fMRI data. We found that 1D CNNs were the best-performing models for fMRI time series classification in our application. We conducted several ablation experiments to further investigate the inherent characteristics of the fMRI time series. Overall, our results suggested that spatial organization and cross-ROI dependencies contain the most discriminative information for classification. We also identified functional networks relevant to sex classification: association networks were the most discriminative, whereas unimodal sensory–motor systems were the least. Although our conclusions are based on a single cohort, parcellation, and architectural choices, they establish baselines and provide insights for designing deep learning architectures for fMRI time series analysis, with potential extension to other phenotypes.

\newpage
\section*{Acknowledgments} Data were provided (in part) by the Human Connectome
Project, WU-Minn Consortium (Principal Investigators:
David Van Essen and Kamil Ugurbil; 1U54MH091657)
funded by the 16 NIH institutes and Centers that support the
NIH blueprint for Neuroscience Research; and by the
McDonnell Center for Systems Neuroscience at Washington
University.


\section*{Conflict of interest} No competing interests exist.


\bibliographystyle{unsrt}  
\bibliography{references}  
\end{document}